\documentclass[aps,prd,twocolumn]{revtex4}

\usepackage{epsfig}
\usepackage{longtable}

\begin{document}

\title{$Z'$ Discovery Reach at Future Hadron Colliders: A Snowmass White Paper} 
\author{ Stephen Godfrey$^1$\footnote{Email: godfrey@physics.carleton.ca} and 
Travis Martin$^2$\footnote{Email: tmartin@triumf.ca}}
\affiliation{
$^1$ Ottawa-Carleton Institute for Physics, 
Department of Physics, Carleton University, Ottawa, Canada K1S 5B6 
\\ 
$^2$ TRIUMF, 4004 Wesbrook Mall, Vancouver, Canada V6T 2A3}

\date{\today}

\begin{abstract}
Extra neutral gauge bosons are a feature of many models of physics beyond the standard model (BSM) and
their discovery could possibly be the first evidence for new physics.  In this Snowmass white paper 
we compare the discovery reach of the high energy hadron colliders considered by the Snowmass study for
a broad range of BSM models.  
It is expected that the LHC should be able to see evidence for a $Z^\prime$ arising from a 
large variety of BSM models up to a mass of $\sim 5$~TeV when the LHC reaches its design energy and luminosity, 
and up to $\sim 6$~TeV with the high luminosity upgrade.
Further into the future, the high energy LHC would 
substantially extend this reach to $\sim 11$~TeV, while the 100~TeV VHE-LHC could see evidence for $Z'$'s 
up to $\sim 30$~TeV.
\end{abstract}
\maketitle

\section{Introduction}

Extra gauge bosons, including $Z'$'s and $W'$'s,  
are a feature of many models of physics beyond the SM
\cite{Cvetic:1995zs,Rizzo:2006nw,Leike:1998wr,Langacker:2008yv,Hewett:1988xc}. 
Examples of such models are Grand Unified theories based on groups
such as $SO(10)$ or $E_6$ \cite{Hewett:1988xc},  
Left-Right symmetric models \cite{lrmodels}, Little Higgs models 
\cite{Perelstein:2005ka,ArkaniHamed:2002qy,Schmaltz:2004de,Han:2003wu},
and Technicolour models \cite{Chivukula:1994mn,Simmons:1996ws,Hill:1994hp,Lane:1995gw}
 to name a few.  In addition, resonances that arise as Kaluza-Klein excitations in theories
of finite size extra dimensions \cite{Hewett:2002hv}
would also appear as new gauge bosons in high energy experiments.
It is therefore quite possible that the discovery of a new gauge boson could be one of the
first pieces of evidence for physics beyond the SM.  Evidence for extra gauge bosons can take two forms:
either from direct production, or indirectly via deviations from standard model predictions.
The first approach is limited by the kinematic reach of direct production at hadron colliders or $e^+e^-$ 
colliders.  However, given that the current direct limits of $Z'$'s are $\sim 3$~TeV, it is clear that $Z'$'s 
will not be produced directly at any of the high energy lepton colliders envisaged for the foreseeable
future.

Currently, the highest mass bounds on most extra neutral gauge bosons are obtained by searches at
the Large Hadron Collider by the ATLAS and CMS experiments.  Recent results 
based on dilepton resonance searches in $\mu^+\mu^-$ and $e^+e^-$ final states
use data from the  
7 TeV proton collisions collected in 2011 and the more recent 8 TeV data collected in 2012.  
ATLAS \cite{ATLAS-ZP-2013} obtains exclusion limits at 95\% C.L. of
$M(Z^\prime_{\rm SSM})>2.86$~TeV, $M(Z^\prime_\eta)>2.44$~TeV, 
$M(Z^\prime_\chi)>2.54$~TeV and $M(Z^\prime_\psi)> 2.38$~TeV 
from 8 TeV collisions with 20 fb$^{-1}$ integrated luminosity, 
while CMS \cite{CMS-ZP-2012} obtains 95\% C.L. exclusion
limits of $M(Z^\prime_{\rm SSM})>2.96$~TeV and $M(Z^\prime_\psi)> 2.60$~TeV
from 8 TeV collisions using $\sim 20$ fb$^{-1}$ of luminosity luminosity.
For these values, SSM refers to the sequential standard model which has the same fermion 
couplings as the Standard Model
and is often used as a reference when comparing constraints from different measurements.
The labels $\eta$, $\chi$, and $\psi$ refer to $Z'$'s arising from 
different symmetry breaking scenarios of the $E_6$ group \cite{Hewett:1988xc}.

\section{$Z'$ Production at Hadron Colliders}

Hadron colliders can produce a $Z'$ boson via Drell-Yan production 
\cite{Rizzo:2006nw,hep-ph/0201093,Godfrey:1987qz,Dittmar:2003ir,Diener:2010sy}, 
which would then be observed 
in the invariant mass distribution of the pair produced final state particles. For most models with a $Z^\prime$, 
the cleanest final state is dileptons, both muons and electrons, due to low backgrounds and clean 
identification.  A very small number of dilepton events clustered in one or two bins of the invariant mass 
distribution would be taken as an obvious signal for new physics.  To quantify this, we consider two opposite 
sign leptons and impose kinematic cuts of $|\eta_l|<2.5$ and $p_{T_l}>20$~GeV to reflect detector 
acceptance. For the discovery limits, we assumed a criteria of 5 isolated dimuon pair events, with a 
signal-over-background ratio of at least 5, within $\pm 3\Gamma_{Z^\prime}$ of the resonance peak.

\begin{figure}[t]
\begin{center}
\centerline{\epsfig{file=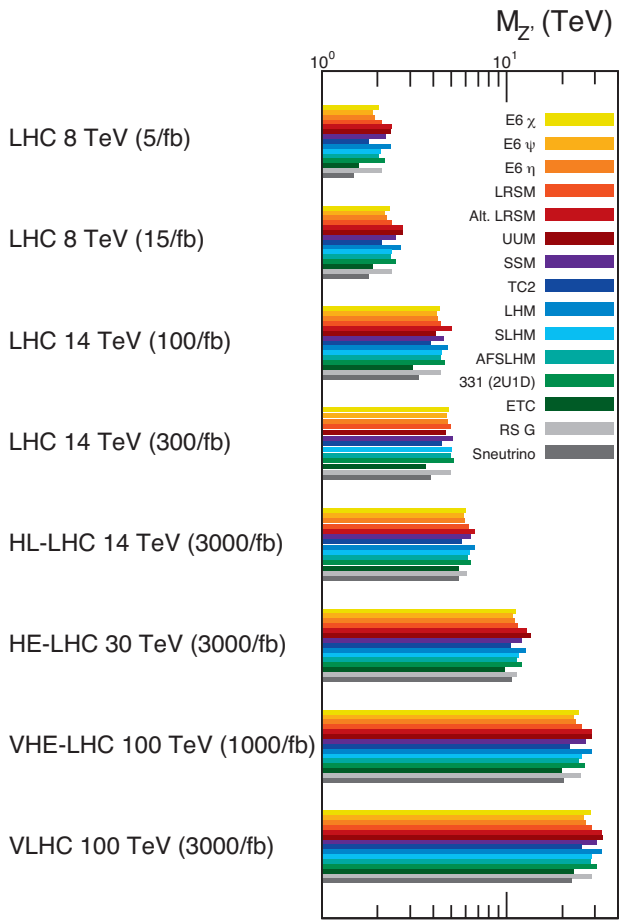,width=3.2in,clip=}}
\end{center}
\caption{$Z'$ discovery reach at high energy hadron colliders. }
\label{Fig1}
\end{figure}

In Fig.~\ref{Fig1} we show the discovery limits for the various models for several hadron collider
benchmark energies and luminosities. When the LHC reaches it's design energy and luminosity it should 
be able to see evidence for $Z'$'s up to $\sim 5$~TeV for a large variety of BSM models
\cite{Godfrey:1994qk,Rizzo:1996ce,hep-ph/0201093,Diener:2010sy,Erler:2011ud}, and the high luminosity 
upgrade (HL-LHC) will extend this reach up to $\sim 6$~TeV.  The high energy LHC (HE-LHC) 
would substantially extend this reach to $\sim 11$~TeV while the 100~TeV very high energy LHC (VHE-LHC) 
could see evidence for 
$Z'$'s up to $\sim 30$~TeV.
For these future searches, the di-electron final state will be the most important as the electromagnetic 
calorimeters will still be able to measure the electron energy accurately, while the muon energy, measured 
from the curvature of the track, will not be easily measurable with the existing magnetic field for muons with 
transverse momenta much above 1 TeV \cite{Aad:2009wy}.

For models with a leptophobic $Z^\prime$ and $W^\prime$, the dominant decay mode for the gauge bosons are to
diquarks and leads to a resonance in the dijet invariant mass distribution.
Searches have been performed
that require two well-separated jets with high transverse momentum.  The CMS collaboration
excludes the existence of a SSM $Z'$ boson with mass between 1.20 and 1.68 TeV at 95\% C.L. 
and a SSM $W'$ with mass between 1.20 and 2.29~TeV  using the $\sqrt{s}=8$~TeV, 
${\cal L}_{\rm int}=20.0$ fb$^{-1}$ dataset \cite{CMS:2012eza}.  The CMS 
collaboration have also developed a dedicated search for $b\bar{b}$ resonances and excluded the
existence of a SSM $Z'$ boson within the range of 1.20 to 1.68 TeV at 95\% C.L. in the $b\bar{b}$ 
channel \cite{CMS:2012dxa}. For 
models with larger branching fractions to $b$-quarks the limit improves considerably, excluding 
a larger mass range.

To search for evidence of models with generation dependent couplings, such as technicolor models, 
both CMS and ATLAS have performed searches for $Z^\prime$ resonances in the $t\bar{t}$ final state. 
The CMS searches in the dilepton plus jets final state\cite{Chatrchyan:2012yca} have excluded masses 
below 1.3~TeV (1.9~TeV) for a $Z^\prime$ width of $\Gamma_{Z^\prime}/M_{Z^\prime} = 0.012$ (0.10), 
while the search in the lepton plus jets final state\cite{Chatrchyan:2012cx} have excluded masses 
below 1.49~TeV (2.04~TeV). Similarly, ATLAS has excluded a $Z^\prime$ resonance decaying to $t\bar{t}$ 
in topcolor assisted technicolor models within the mass range of $0.5 < M_{Z^\prime} < 1.8$~TeV at 95\% C.L. 
in the single lepton plus jets final state, using 14 fb$^{-1}$
 integrated luminosity \cite{ATLAS-2012-052}.   
It is more difficult to project limits in the $t\bar{t}$ and
$b\bar{b}$ channels to higher energies because of the added complications in dealing with hadronic final states  
so this topic is left for a future study.

\section{Effects of $Z'$'s at High Energy $e^+ e^-$ Colliders}

The other way to search for evidence for extra neutral gauge bosons is to look for deviations from
the standard model due to interference and virtual effects.  In particular, high energy $e^+e^-$ colliders will 
be sensitive to new gauge bosons with $M_{Z'} \gg \sqrt{s}$.  We mention results 
for the sensitivity of high energy
$e^+e^-$ colliders to $Z'$'s for comparison purposes and refer the interested reader to the existing 
literature \cite{Godfrey:1994qk,Rizzo:1996ce,Godfrey:1996uq,Osland:2009dp} including another contribution 
to the 2013 Snowmass study \cite{Han:2013mra}.  

In $e^+e^-$ collisions below the on-shell production threshold, 
extra gauge bosons manifest themselves as deviations
from SM predictions due to interference between the new physics and the SM $\gamma /Z^0$ contributions.  
$e^+e^-\to f\bar{f}$ reactions are characterized by relatively clean, simple final states
where $f$ could be leptons ($e$, $\mu$, $\tau$) or  quarks ($u$, $d$, $s$, $c$, $b$, $t$), for both 
polarized and unpolarized $e^\pm$. 
The basic $e^+e^-\to f\bar{f}$ processes can be parametrized in terms of four 
helicity amplitudes which 
can be determined by measuring various observables: the leptonic cross section,
$\sigma(e^+e^-\to \mu^+\mu^-)$,
the ratio of the hadronic to the QED point cross section $R^{\rm had}=\sigma^{\rm had}/\sigma_0$,
the leptonic forward-backward asymmetry, $A^\ell_{\rm FB}$, the leptonic  longitudinal asymmetry, 
$A^\ell_{\rm LR}$, the hadronic longitudinal asymmetry, $A^{\rm had}_{\rm LR}$, 
the forward-backward asymmetry for specific quark or lepton 
flavours, $A^f_{\rm FB}$, the $\tau$ polarization asymmetry, 
$A_{pol}^\tau$, and the polarized forward-backward asymmetry for 
specific fermion flavours, $A^f_{\rm FB}({\rm pol})$ \cite{Godfrey:1996uq}.
The indices $f=\ell, \; q$, $\ell =(e,\mu,\tau)$, $q=(c, \; b)$, 
and $had=$`sum over all hadrons' indicate the final state fermions.
Precision measurements of these observables 
for various final states ($\mu^+\mu^-$, $b\bar{b}$, $t\bar{t}$) can 
be sensitive to extra gauge boson masses that by far exceed the
direct search limits that are expected at the LHC
\cite{Godfrey:1994qk,hep-ph/0201093,Godfrey:1996uq,Osland:2009dp,Han:2013mra}.  
Further,  precision measurements of cross sections
to different final state fermions using polarized beams can be used to constrain the gauge boson
couplings and help distinguish the underlying theory
\cite{Osland:2009dp,Han:2013mra,Leike:1996qq,Riemann:2001,Godfrey:2005pm,Linssen:2012hp,Battaglia:2012ez}.  
A deviation for one observable is always possible as 
a statistical fluctuation and different observables have different sensitivities to
different models (or more accurately to different couplings).  As a consequence, 
a more robust strategy is to combine many observables to obtain a $\chi^2$ figure of merit.

The ILC sensitivity to $Z'$'s is based on high statistics precision cross section measurements 
so that the reach will depend on the integrated luminosity.  
For many models a 500~GeV $e^+e^-$ 
collider with as little as 50~fb$^{-1}$ integrated luminosity would see the effects of a $Z'$ with 
masses as high as $\sim 5$~TeV \cite{Godfrey:1994qk}.  
Recent studies \cite{Osland:2009dp,Han:2013mra}
find that a 500~GeV ILC with 500~fb$^{-1}$ and a 1~TeV ILC with 1~ab$^{-1}$
can see evidence or rule out a $Z'$ with masses that can exceed  $\sim 7$ and $\sim 12$~TeV respectively,
for many models.  
These results also
consider various polarizations for the $e^-$ and $e^+$ beams and show that beam polarization 
will increase the potential reach of the ILC.
It should be noted
that in the case of $e^+e^-$ colliders the exclusion limits are very sensitive to the specific model in
contrast to hadron colliders where the exclusion limits are are far less model dependent.

\section{Final Comments}

We presented the discovery reach for extra gauge bosons that are possible for various high energy hadron colliders
that have been considered for the Snowmass study.  
The LHC  should 
be able to see evidence for $Z'$'s up to $\sim 5$~TeV for a large variety of BSM models, the HL-LHC 
will extend this reach up to $\sim 6$~TeV,  the 30~TeV HE-LHC 
 to $\sim 11$~TeV  and  the 100~TeV VHE-LHC up to $\sim 30$~TeV.  In comparison, a 500~GeV, 500~fb$^{-1}$ $e^+e^-$ 
collider would be sensitive to $Z'$'s $\sim 6$~TeV comparable to the HL-LHC and 
a 1~TeV, 1~ab$^{-1}$ $e^+e^-$ collider would be sensitive to $Z'$'s $\sim 12$~TeV comparable to the HE-LHC.  
We did not discuss the issue of $Z'$ identification if a $Z'$ were discovered.  If a $Z'$ were discovered at
a hadron collider, precision measurements at a high energy $e^+e^-$ collider would give valuable 
information on it's couplings which could not be obtained at hadron colliders on the same time scale.

\acknowledgments

This research was supported in part 
the Natural Sciences and Engineering Research Council of Canada.


\end{document}